\documentclass[12pt,showpacs,amsmath,amssymb]{revtex4}
\usepackage{bm}
\usepackage{dcolumn}
\usepackage{graphicx}
\usepackage{epsf}
\usepackage{epsfig}
\begin{document}

\title{Explicit Fermi Coordinates and Tidal Dynamics in de Sitter and G\"odel
Spacetimes}

\author{C. Chicone}

\affiliation{Department of Mathematics, University of Missouri-Columbia, Columbia,
Missouri 65211, USA }

\author{B. Mashhoon}

\affiliation{Department of Physics and Astronomy, University of Missouri-Columbia,
Columbia, Missouri 65211, USA}

\begin{abstract}Fermi coordinates are directly constructed in de Sitter and
G\"odel spacetimes and the corresponding exact coordinate transformations are
given explicitly. The quasi-inertial Fermi coordinates are then employed to
discuss the dynamics of a free test particle in these spacetimes and the
results are compared to the corresponding generalized Jacobi equations that
contain only the lowest-order tidal terms. The domain of validity of the
generalized Jacobi equation is thus examined in these cases. Furthermore, the difficulty of constructing explicit Fermi coordinates in black-hole spacetimes is demonstrated.
\end{abstract}
\pacs  {04.20.Cv}

\maketitle
\section{introduction}

To interpret measurements in a gravitational field, access to locally
inertial coordinates is indispensable. At each event in spacetime, the
spacetime is locally flat; therefore, it is possible to introduce
Riemann normal coordinates that constitute a geodesic system of
coordinates that is inertial at the event under consideration. The
Riemann normal coordinates are in general admissible only in a certain
region around the event such that every point in this region can be
connected to the event by a unique geodesic.

In general, the physical interpretation of measurements by a free
observer necessitates a continuous locally inertial system along the
worldline of the observer. This can be achieved by Fermi coordinates,
namely, a normal geodesic coordinate system in a cylindrical region
about the worldline of the observer. Fermi normal coordinates~\cite{1}
are the natural extension of Riemann normal coordinates and play a
basic role in general relativity theory~\cite{2}. The notion of Fermi coordinates was first implicitly introduced in Ref.~\cite{1}; a recent
commentary on Fermi's pioneering work is contained in Ref.~\cite{3}.
\begin{figure}
\centerline{\psfig{file=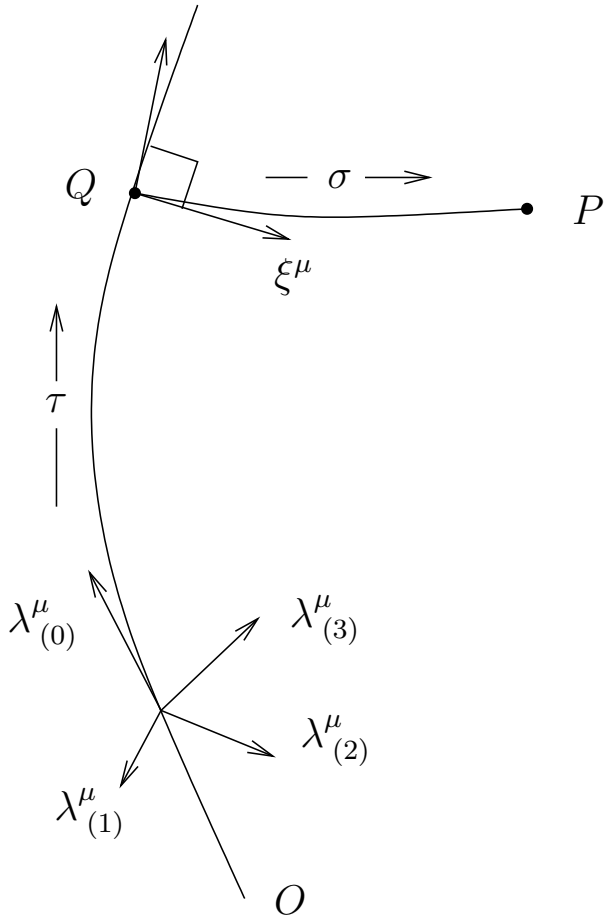, width=20pc}}
\caption{Schematic construction of Fermi coordinates at $P$.
\label{fig:1}}
\end{figure}

Imagine a spacetime region with coordinates $x^\mu =(t,x^i)$ and a
reference observer $O$ following a worldline $(\bar{t},\bar{x}^i)$. We
use gravitational units such that $c=G=1$ throughout; moreover, we assume that the spacetime metric has signature $+2$. The observer
carries an orthonormal parallel-propagated tetrad frame $\lambda^\mu
_{\;\;(\alpha)}$ along its path such that
$\lambda^\mu_{\;\;(0)}=d\bar{x}^\mu /d\tau$, where $\tau$ is the
proper time along the worldline of $O$. Thus $\lambda^\mu_{\;\;(0)}$
is the timelike unit vector that is tangent to the worldline of $O$
and acts as its local temporal axis. Moreover,
$\lambda^\mu_{\;\;(i)}$, $i=1,2,3$, are orthogonal spacelike unit gyro
axes that form the local spatial frame of the observer. At each event
$Q(\tau)$ along the worldline, consider the class of spacelike
geodesics orthogonal to the worldline; these form a local
hypersurface. Let $P$ be an event with coordinates $x^\mu$ on this
hypersurface and consider the {\it unique} spacelike geodesic segment
from $Q$ to $P$. We define the Fermi coordinates of $P$ to be $X^\mu
=(T,X^i)$, where
\begin{equation}
\label{eq1} T=\tau ,\quad X^i=\sigma \xi ^\mu \lambda_\mu^{\;\; (i)}.
\end{equation}
Here $\xi^\mu$ is the unit vector tangent to the spacelike geodesic
segment at $Q$ and $\sigma$ is the proper length of this segment from
$Q$ to $P$ as in Figure 1. Thus the reference observer $O$ is always
at the spatial origin of Fermi coordinates.

The Fermi coordinates are in general admissible in the sense of
Lichnerowicz~\cite{4} in a finite cylindrical region about the
worldline of $O$. The spacetime metric in Fermi coordinates is given
by
\begin{align}
\label{eq2} g_{00} &= -1- {^F\! R_{0i0j}} (T)X^iX^j+\cdots ,\\
\label{eq3} g_{0i} &= -\frac{2}{3} {^F\! R_{0jik}}(T)X^jX^k+\cdots ,\\
\label{eq4} g_{ij} &= \delta_{ij}-\frac{1}{3}
      {^F\! R_{ikjl}}(T)X^kX^l+\cdots ,
\end{align}
where
\begin{equation}
\label{eq5} {^F\! R_{\alpha\beta\gamma \delta}} (T)=R_{\mu\nu\rho \sigma}
  \lambda^\mu_{\;\;(\alpha)} \lambda^\nu_{\;\; (\beta)}
  \lambda^\rho_{\;\;(\gamma)} \lambda^\sigma _{\;\;(\delta)} 
\end{equation}
is the projection of the Riemann tensor on the observer's tetrad along
the reference trajectory. The Taylor series in
Eqs. \eqref{eq2}--\eqref{eq4} can be expressed as
$g_{\mu\nu}=\eta_{\mu\nu} +h_{\mu\nu}(X)$, where $h_{\mu\nu}(X)$ is a
perturbation that can be expressed as a series expansion in powers of
the spatial distance away from the reference trajectory. The nature of
the infinite series in Eqs.~\eqref{eq2}--\eqref{eq4} has been discussed
by a number of authors. The second-order terms given explicitly in
Eqs.~\eqref{eq2}--\eqref{eq4} were worked out in Refs.~\cite{5,6}. The third-order terms were first given in Ref.~\cite{7} and the fourth-order terms in Ref.~\cite{8}; moreover, these higher-order terms have been recently discussed in
Ref.~\cite{9}. In the weak-field limit, the infinite series in
Eqs. \eqref{eq2}--\eqref{eq4} have been studied in Ref.~\cite{10}.

Consider the motion of a free particle in the Fermi coordinate
system. The geodesic equation of motion 

\begin{equation}
\label{eq6} \frac{dU^\mu}{ds}+\Gamma^\mu_{\rho\sigma} U^\rho U^\sigma
=0,
\end{equation}
where $U^\mu=dX^\mu/ds$, can be written in terms of the modified
Lorentz factor $\Gamma=dT/ds$ and $\mathbf{V}=d\mathbf{X} /dT$ based
on the decomposition $U^\mu =\Gamma (1,\mathbf{V})$. On the other hand,
the timelike condition $U^\mu U_\mu =-1$ implies that

\begin{equation}
\label{eq7} -\frac{1}{\Gamma^2} =g_{00}+2g_{0i} V^i+g_{ij}V^iV^j.
\end{equation}
The equation for $\mathbf{V}(T)$ is the reduced geodesic equation
given by

\begin{equation}
\label{eq8} \frac{dV^i}{dT}+ (\Gamma^i_{\alpha\beta}-\Gamma^0_{\alpha
  \beta}V^i)\frac{dX^\alpha}{dT} \frac{dX^\beta}{dT}=0.
\end{equation}
It is useful to consider the class of static observers in the Fermi
coordinate system. These observers are generally accelerated with
four-velocity $U^\mu_S=(-g_{00})^{-\frac{1}{2}}\delta^\mu_{\;\;0}$ in
Fermi coordinates. The Lorentz factor of the test particle with four-velocity
$U^\mu=\Gamma (1,\mathbf{V})$ with respect to the static observers is

\begin{equation}
\label{eq9} \Gamma_S=\left(\sqrt{-g_{00}}-\frac{g_{0i}V^i}{\sqrt{-g_{00}}}\right) \Gamma .
\end{equation}

In this paper, we are interested in the comparison between the
consequences of these equations in the case of exact Fermi
coordinates---given explicitly in de Sitter and G\"odel
spacetimes---with equations based on the lowest-order tidal terms
given  explicitly in Eqs.~\eqref{eq2}--\eqref{eq4}. In the latter case,
Eq.~\eqref{eq7} implies that
\begin{equation}\begin{split}
\label{eq10}\frac{1}{\Gamma^2}&=1-V^2+ {^F\! R_{0i0j}}X^iX^j+\frac{4}{3}
  {^F\! R_{0jik} }X^jV^iX^k\\
&\quad + \frac{1}{3} {^F\! R_{ikjl}}V^iX^kV^jX^l
\end{split}\end{equation}
and the reduced equation of motion is the generalized Jacobi
equation~\cite{11}
\begin{equation}\begin{split} 
\label{eq11} &\frac{d^2X^i}{dT^2} +{^F\! R_{0i0j}}X^j+2 {^F\! R_{ikj0}}V^kX^j\\
&\quad +\frac{2}{3} (3 {^F\! R_{0kj0}} V^iV^k +
    {^F\! R_{ikjl}}V^kV^l     +{^F\! R_{0kjl}}V^iV^kV^l)X^j=0.
\end{split}\end{equation}
This equation is expected to be valid for $|\mathbf{X}|$ sufficiently small
compared to a certain radius of curvature of spacetime. 
We assume the following initial conditions for Eqs.~\eqref{eq8} and~\eqref{eq11} throughout
this paper: At $T = 0$, $\mathbf{X} = 0$ and $\mathbf{V} =\mathbf{V}_0$ such that $|\mathbf{V}_0|< 1$. Let us note that for $T > 0$ and away from the origin of Fermi coordinates, $|\mathbf{V}|$ could then in principle exceed unity along a timelike geodesic worldline.

In some situations---for instance, along the axis of rotational
symmetry of a Kerr black hole---one-dimensional motion is allowed. In
general, the symmetries of the Riemann tensor imply that for motion
along the $Z$ direction, say, Eq.~\eqref{eq11} reduces to
\begin{equation}
\label{eq12} \frac{d^2Z}{dT^2} +\kappa (1-2\dot{Z}^2)Z=0,
\end{equation}
where $\kappa(T)= {^F\!R_{TZTZ}} (T)$ and $\dot{Z}=dZ/dT$. This equation approximates the reduced geodesic equation for $|Z|$
sufficiently small compared to $|\kappa|^{-1/2}$.  In
Eq.~\eqref{eq12}, the critical speed of $1/\sqrt{2}\approx 0.7$ should be noted:
Motion at the critical speed is uniform, i.e. $Z=\text{ constant }\pm
T/\sqrt{2}$, and the character of the motion changes depending on
whether the initial relative speed is above or below the critical
speed. The critical speed in gravitational motion is briefly reviewed in
Appendix A.

In a recent series of papers, we have investigated the general
equation of motion of a particle in Fermi coordinates using the generalized Jacobi equation~\cite{11}. The
results are particularly interesting if the speed of the test particle
relative to the reference particle exceeds the critical speed given by
$V_c:=1/\sqrt{2}$. The astrophysical implications of these results have
been worked out in Refs.~\cite{12,13}. This general approach has been
extended to the motion of charged particles in Ref.~\cite{14}. A major
shortcoming of these studies is that only the first few terms of the
series in Eqs.~\eqref{eq2}--\eqref{eq4} have been taken into account.  It is therefore important to know the domain of validity of these
astrophysically significant results.  In
fact, as explicitly demonstrated in Appendix B, the construction of the exact Fermi coordinates in black-hole
spacetimes is a daunting task. Moreover, the transformation
$x^\mu \mapsto X^\mu$ remains implicit as well as approximate in these
studies~\cite{11,12,13}. For recent efforts to construct such approximate
transformations explicitly, see Ref.~\cite{15}.

To avoid the difficulties encountered in black-hole spacetimes ( cf.~Appendix B), we turn to spacetimes with more symmetries. It is clear that
explicit Fermi coordinates can only be constructed in rather special
circumstances; however, the generalized Jacobi equation can be employed in
general. What is the extent of agreement between these approaches? To answer
this question, we choose two spatially homogeneous spacetimes: de Sitter's
spacetime with ten Killing vector fields and G\"odel's spacetime with five
Killing vector fields.

The main purpose of the present paper is to study the infinite series
in Eqs.~\eqref{eq2}--\eqref{eq4} and the exact transformation 
$x^\mu \mapsto X^\mu$ to Fermi coordinates explicitly in de Sitter and
G\"odel spacetimes. The motion of a test particle is then studied using the exact Fermi metric and the results are compared to the implications of the generalized
Jacobi equation ~\eqref{eq12}, where in de Sitter (G\"odel) spacetime $\kappa$ is a
negative (positive) constant. Eq.~\eqref{eq12}  can be integrated in these cases;
the behavior of the solutions have been studied in Ref.~\cite{11}. The comparison
of the exact geodesic equation of motion in Fermi coordinates with the generalized
Jacobi equation makes it in principle possible to determine the extent of validity of the
latter equation in these cases; moreover, a general mathematical treatment of such a comparison is presented in Appendix C. 

\section{Fermi coordinates in de Sitter Spacetime\label{s2}}

Consider de Sitter's metric in the (inflationary-model) form
\begin{equation}
\label{eq13} ds^2=-dt^2+\mathcal{A}^2(t)\delta_{ij}dx^idx^j,
\end{equation}
where $\mathcal{A}(t)$ is given by
\begin{equation}
\label{eq14} \mathcal{A}(t)=e^{Ht}
\end{equation}
and $H>0$ is a (``Hubble'') constant. This metric satisfies Einstein's
matter-free equations such that $H^2=\Lambda /3$, where $\Lambda >0$
is the cosmological constant. The spacetime region
$x^\mu=(t,\mathbf{x})$ described by equation~\eqref{eq1} is only part
of de Sitter's spacetime of constant positive curvature; a detailed historical description of
spacetimes of constant curvature is contained in Ref.~\cite{16}.

The geodesics of the metric~\eqref{eq13} are obtained from the extrema
of $\int ds$. It is a simple consequence of this fact that
$\mathcal{A}^2\,dx^i/ds$, $i=1,2,3$, are constants along the geodesics. Let us
imagine a class of reference observers that are fixed in space,
i.e. they have constant spatial coordinates $x^i$, $i=1,2,3$. It turns
out that these are free observers each with an orthonormal tetrad
frame $\lambda^\mu_{\;\;(\alpha)}$ such that
\begin{equation}
\label{eq15} \lambda^\mu_{\;\; (0)}=\delta^\mu_{\;\;0},\quad
\lambda^\mu_{\;\;(i)}=\frac{1}{\mathcal{A}(t)}\delta^\mu_{\;\; i},
\end{equation}
where $t=\tau$ is the proper time along the worldline of the
observer. Moreover, it is simple to verify that this tetrad frame is
parallel transported along the geodesic worldline of the observer.

The projection of the Riemann tensor on the tetrad of a reference
observer can be expressed as a $6\times 6$ matrix
$\mathcal{R}=(\mathcal{R}_{AB})$, where $A$ and $B$ are elements of
the set $\{ 01,02,03,23,31,12\}$. We find that
\begin{equation}
\label{eq16} \mathcal{R}=-H^2\begin{bmatrix}I & \quad 0\\ 0 &\quad
-I\end{bmatrix},
\end{equation}
where $I$ is the $3\times 3$ unit matrix.

We need to find the general solution for {\it spacelike} geodesics. To
this end, the  spacelike geodesic equation in this case reduces to
\begin{align}
\label{eq17} \frac{dx^i}{d\sigma} &= \frac{C_i}{\mathcal{A}^2(t)} ,\\
\label{eq18} \left( \frac{dt}{d\sigma} \right)^2 &= \frac{C^2}{\mathcal{A}^2(t)}
-1,
\end{align}
where $\sigma$ is the proper length along a spacelike geodesic,  $C_i$,
$i=1,2,3$,  are constants and
\begin{equation}
\label{eq19} C^2=\sum_i C_i^2.
\end{equation}
The general solution of Eq.~\eqref{eq18} is
\begin{equation}
\label{eq20} e^{Ht}=C\cos (H\sigma +\eta ),
\end{equation}
where $\eta$ is a constant. It follows from Eq.~\eqref{eq20} that
Eq.~\eqref{eq17} can be solved exactly and the result is
\begin{equation}
\label{eq21} x^i=\frac{C_i}{HC^2}\tan (H\sigma +\eta) +D_i,
\end{equation}
where $D_i$, $i=1,2,3$, are constants of integration.

To establish a Fermi coordinate system, we need to choose a specific
reference trajectory. For the sake of simplicity and with no loss in
generality, we choose the reference observer
$O:(\bar{t},\mathbf{\bar{x}})=(\tau ,\mathbf{0})$. In the construction
of Fermi coordinates (see Figure~1), the spacelike geodesic
segment from $Q$ to $P$ is given by equations \eqref{eq20} and
\eqref{eq21} such that $\sigma =0$ at $Q$. Therefore,
\begin{equation}
\label{eq22} e^{H\tau}=C\cos \eta ,\quad \frac{C_i}{HC^2} \tan \eta
+D_i=0.
\end{equation}
Moreover, $\xi^\mu$ is the vector tangent to the spacelike geodesic
segment at $Q$ and is given in $(t,\mathbf{x})$ coordinates by
\begin{equation}
\label{eq23} \xi ^\mu =\left(-\tan \eta ,
\frac{C_i}{C^2\cos^2\eta}\right).
\end{equation}
It follows from $\xi_\mu \lambda^\mu_{\;\;(0)}=0$ that $\tan \eta =0$
and hence $D_i=0$, $i=1,2,3$, from Eq.~\eqref{eq22}. We choose $\eta
=0$ for simplicity; hence, $C=\exp (H\tau )$. Finally, we compute the
spatial Fermi coordinates of $P$ using Eq.~\eqref{eq1}. It follows
that event $P$ has Fermi coordinates
\begin{equation}
\label{eq24} T=\tau,\quad \mathbf{X} =\sigma e^{-H\tau}\mathbf{C}.
\end{equation}
It is useful to recognize that $\sigma$ is the radial Fermi coordinate
$R$,
\begin{equation}
\label{eq25} R=\sqrt{\delta_{ij}X^iX^j},
\end{equation}
so that we find from Eqs.~\eqref{eq20} and \eqref{eq21} that the
transformation $(t,\mathbf{x})\mapsto (T,\mathbf{X})$ is given by
\begin{align}
\label{eq26} e^{Ht}&=e^{HT}\cos (HR),\\
\label{eq27} \mathbf{x}&=e^{-HT}\, \frac{\tan(HR)}{HR}\,\mathbf{X} .
\end{align}
De Sitter's metric in Fermi coordinates is $ds^2=g_{\mu\nu} dX^\mu
dX^\nu$, where
\begin{align}
\label{eq28} g_{00}&= -\cos ^2 (HR),\quad g_{0i}=0,\\
\label{eq29} g_{ij}&=\frac{X^iX^j}{R^2} +\frac{\sin^2(HR)}{H^2R^2}
\left( \delta_{ij}-\frac{X^iX^j}{R^2} \right).
\end{align}
The Fermi coordinates cover a \textit{static} spacetime region and are
admissible for $0\leq R<\pi /(2H)$; in fact, $g_{00}=0$ for $HR=\pi
/2$.

The form of the metric \eqref{eq28}--\eqref{eq29} can be simplified if
we introduce Fermi polar coordinates via
\begin{equation}
\label{eq30}X^1=R\sin \theta \cos \phi ,\quad X^2=R\sin \theta \sin\phi,\quad
X^3=R\cos \theta .
\end{equation}
The Fermi metric in these spherical polar coordinates is given by
\begin{equation}
\label{eq31} ds^2=-\cos ^2 (HR)dT^2+dR^2+\frac{1}{H^2} \sin^2 (HR)
(d\theta^2+\sin^2\theta\, d\phi^2).
\end{equation}
It is interesting to note that this form of de Sitter's metric already
appeared in de Sitter's original investigations, namely, in his 1917
paper on the curvature of space (cf. Eq.~(2.5) of Ref.~\cite{16} and
the discussion therein).

Writing $\cos^2(HR)$ and $\sin^2(HR)$ in Eqs.~\eqref{eq28} and
\eqref{eq29} in terms of $\cos (2HR)$ and the subsequent Taylor
expansion of $\cos (2HR)$ about $R=0$ would lead to the standard
series expansion of the elements of the metric tensor in Fermi
coordinates as in Eqs.~\eqref{eq2}--\eqref{eq4}. Such series are
uniformly convergent for all $R$; however, the Fermi coordinates are
admissible only for $0\leq 2HR<\pi$. The hypersurface $R=\pi /(2H)$,
where the timelike Killing vector $\partial_T$ becomes null, is a
static limit surface.

\section{Tidal Dynamics in de Sitter Spacetime\label{s3}}

\begin{figure}
\centerline{\psfig{file=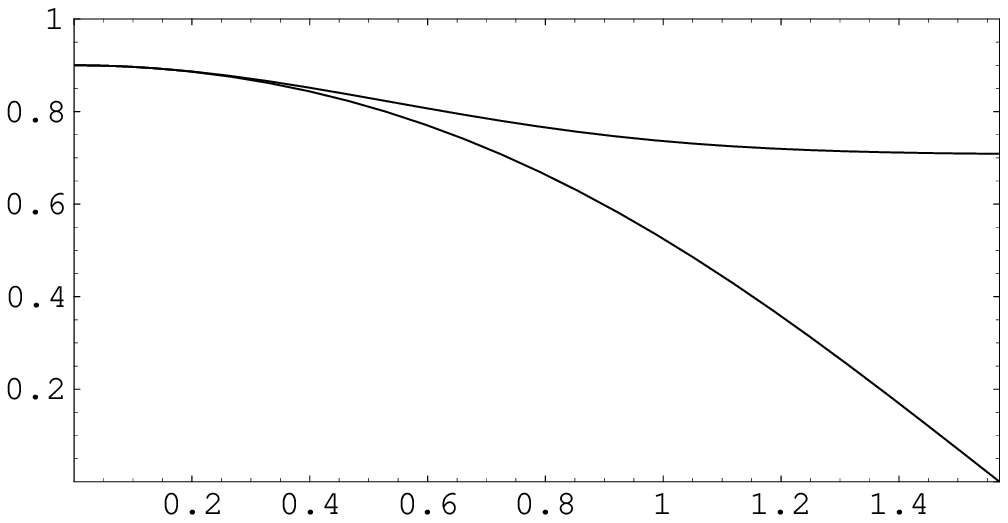, width=15pc}}
\centerline{\psfig{file=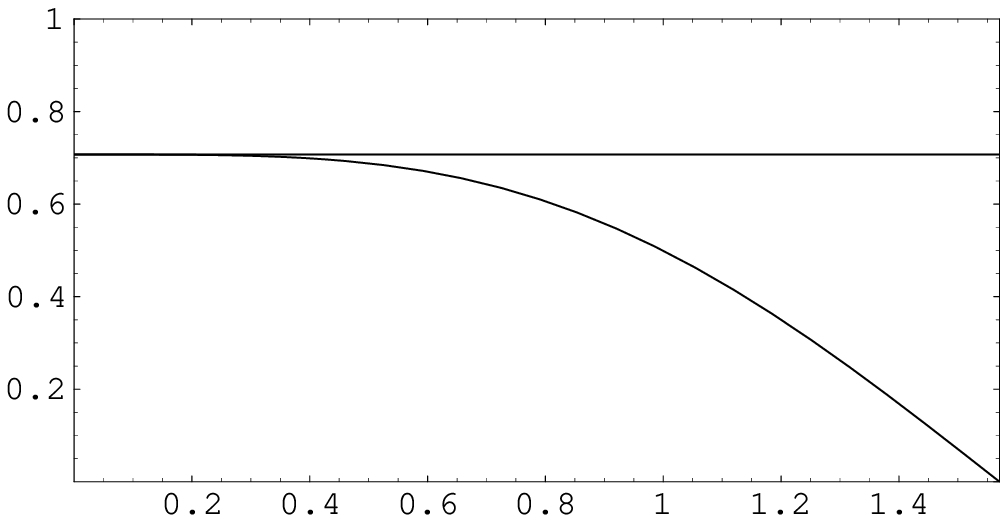, width=15pc}}
\centerline{\psfig{file=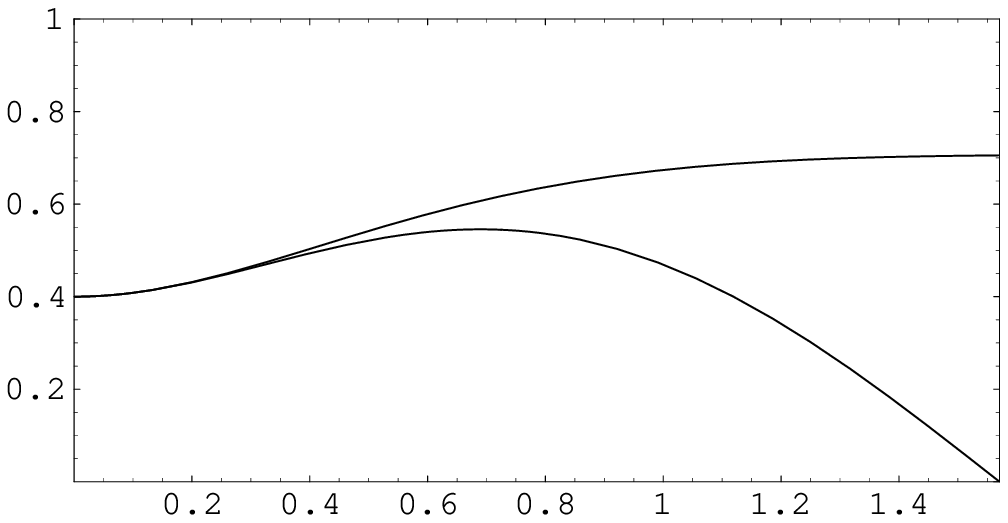, width=15pc}}
\caption{Plot of $\dot R$ versus $HR$, $0\le HR\le \pi/2 $, for $V_0=0.4$, $1/\sqrt{2}$ and $0.9$. In each panel, the lower (upper) curve represents the motion according to the exact (approximate) equation of motion in Fermi coordinates. In the exact case, $\dot R$ vanishes at $HR=\pi/2$. 
\label{fig:2}}
\end{figure}

Let us now consider the radial motion of a test particle away from the
reference observer $O$ at $R=0$ in the spherical Fermi
coordinates~\eqref{eq30}. We concentrate on the general reduced
geodesic equation~\eqref{eq8}. In the case of radial motion in the
spherically symmetric spacetime region of Eq.~\eqref{eq31}, the only relevant
nonzero connection coefficients are
\begin{equation}
\label{eq32} \Gamma^R_{TT}=-\frac{1}{2}H\sin (2HR),\quad
\Gamma^T_{TR}=\Gamma^T_{RT}=-H\tan (HR).
\end{equation}
Therefore, the radial equation of motion is
\begin{equation}
\label{eq33} \frac{d^2R}{dT^2} -H\tan (HR) \left[
  \cos^2(HR)-2\left(\frac{dR}{dT}\right)^2\right] =0.
\end{equation}
This equation can be easily integrated  once and the result for
$\dot{R}=dR/dT$ is
\begin{equation}
\label{eq34} \dot{R}^2=\cos^2 (HR)-(1-V^2_0)\cos^4(HR),
\end{equation}
where $V_0=\dot{R}(T=0)$ is the initial speed at $R=0$. 
Next, reducing Eq.~\eqref{eq34} to quadrature and using the relation
\begin{equation}\label{neweq35}
\int\frac{dx}{\cos x\, \sqrt{1+\lambda^2\sin^2 x}}=\frac{1}{2\sqrt{1+\lambda^2}}\ln\frac{\sqrt{1+\lambda^2\sin^2 x}\,+\sqrt{1+\lambda^2}\,\sin x}{\sqrt{1+\lambda^2\sin^2 x}\,-\sqrt{1+\lambda^2}\,\sin x}
\end{equation}
with $\lambda^2=-1 +1/V_0^2$, it is possible to show that the general solution
of the radial equation with the specified initial conditions is 
\begin{equation}\label{neweq36}
\tan(HR)=\pm V_0\sinh(HT).
\end{equation}
In general, we require that the Fermi time $T$ increase along the timelike worldline of every observer. Therefore, assuming $V_0>0$, we are only interested in the upper sign in Eq.~\eqref{neweq36}.
The modified Lorentz factors $\Gamma$ and $\Gamma_S$ are given in
this case by
\begin{equation}
\label{eq35} \Gamma =\frac{\Gamma_0}{\cos^2(HR)} ,\quad
\Gamma_S=\frac{\Gamma_0}{\cos (HR)},
\end{equation}
where $\Gamma_0 =(1-V_0^2)^{-\frac{1}{2}}$. The speed of the particle
initially increases, remains constant or decreases depending upon
whether $V_0$ is less than, equal to or greater than the critical
speed $1/\sqrt{2}$; nevertheless, Eq.~\eqref{eq35} indicates that
regardless of the particle's initial speed, as $T\to \infty$ it
monotonically approaches a null geodesic at the static limit surface
$HR=\pi/2$. Moreover, the speed of the particle, as ``seen'' by the
reference observer, is zero at this boundary surface, where
everything, including light, ``appears frozen.''  We note that $\dot R$ has a maximum at $\cos (HR)=\Gamma_0/\sqrt{2}$ for $V_0\le 1/\sqrt{2}$ as illustrated in Figure~\ref{fig:2}. The case of a null
ray follows from Eq.~\eqref{neweq36} with the upper sign for $V_0=1$; in fact, in this case Eq.~\eqref{neweq36}  can also be expressed as $\tanh (HT)=\sin (HR)$.

Let us now determine to what extent these exact results are reflected
in the first-order tidal terms of the generalized Jacobi
Eq.~\eqref{eq11}. For radial motion, we find
\begin{equation}
\label{eq36} \frac{d^2R}{dT^2}=H^2(1-2\dot{R}^2)R,
\end{equation}
which follows as well from Eq.~\eqref{eq33}, since to linear order we
have $\tan(HR)\approx HR$ and $\cos (HR)\approx 1$. Eq.~\eqref{eq36} is thus valid for $R$ sufficiently small compared to $H^{-1}$.  

To study the general behavior of the solution of Eq.~\eqref{eq36}, we note that this equation can be integrated once and the result is
\begin{equation}\label{neweq39}
\dot R^2=V_c^2-(V_c^2-V_0^2) e^{-2 H^2 R^2},
\end{equation}
where $V_c=1/\sqrt{2}$. Let us define a function $\Psi_\pm(x;\nu)$ via the integral
\begin{equation}\label{neweq40}
\Psi_{\pm}(x;\nu):=\int_0^x\frac{dz}{\sqrt{1+\nu e^{\pm z^2}}},
\end{equation}
where $\nu$ is a constant parameter. Near $x=0$, we have the Taylor expansions 
\begin{eqnarray}\label{neweq41}
\Psi_\pm (x;\nu)&=& (\nu+1)^{-1/2}[x\mp\frac{\nu}{\nu+1}\frac{x^3}{3!}+\frac{3\nu(\nu-2)}{(\nu+1)^2}\frac{x^5}{5!}+\cdots],\\
\label{neweq41b} x&=&(\nu+1)^{1/2}[\Psi_\pm \pm\frac{\nu}{3!}\Psi_\pm^3 +\frac{\nu(7\nu+6)}{5!} \Psi_\pm^5+\cdots]. 
\end{eqnarray}
 The general solution of Eq.~\eqref{neweq39} is thus given by
\begin{equation}\label{neweq42}
\Psi_-(\sqrt{2}HR; 2 V_0^2-1)=\pm HT.
\end{equation}
For the physical problem under consideration here only the upper sign in Eq.~\eqref{neweq42} is needed; in this case, the motion can be described in terms of an attractor involving uniform motion at speed $V_c$. Explicitly, it is simple to see via the effective potential in Eq.~\eqref{neweq39} that for $V_0>V_c$, the particle monotonically decelerates and asymptotically (i.e. for $HT\to \infty$ and $HR\to \infty$) approaches the critical speed. For $V_0=V_c$, the particle moves uniformly, while for $V_0<V_c$ the particle monotonically accelerates and asymptotically approaches $V_c$. These results,  together with the  extent of the
(initial) agreement between Eq.~\eqref{neweq42} and the exact result given by 
Eq.~\eqref{neweq36},  are illustrated in Figure~\ref{fig:2}. Analytic estimates for the difference between the exact solution and the approximation  at a given time $T$ can be obtained using Eqs.~\eqref{neweq36} and~\eqref{neweq42}.

\section{Fermi coordinates in G\"odel Spacetime\label{s4}}

Let us next consider the stationary and spatially homogeneous rotating
universe model discovered by G\"odel~\cite{17}. It has been discussed
by a number of authors~\cite{18}. G\"odel's metric can be expressed as
\begin{equation}
\label{eq37} ds^2=-dt^2-2\sqrt{2} U(x) dtdy+dx^2-U^2dy^2+dz^2,
\end{equation}
where
\begin{equation}
\label{eq38} U(x)=e^{\sqrt{2}\Omega x}
\end{equation}
and $\Omega $ is a positive constant. The Ricci curvature for this
spacetime is given by
\begin{equation}
\label{eq39} R_{\mu\nu}=2\Omega^2 u_\mu u_\nu ,
\end{equation}
where $u^\mu =\delta^\mu_{\;\; 0}$ is the four-velocity vector field
for free particles that are at rest in space and coincides with the
timelike Killing vector field $\partial_t$. The source of the G\"odel
gravitational field could be a perfect fluid with velocity $u^\mu$ and
constant density $\mu$ and pressure $\tilde p$ given by $\mu =\tilde p=\Omega^2/(8\pi)$, where
$\Omega \partial_z$ is the vorticity vector of the geodesic worldlines
of the fluid source. Alternatively, the source of the G\"odel field
could be dust of constant density $\Omega^2/(4\pi)$ together with a
cosmological constant $\Lambda=-\Omega^2$~\cite{19}. The investigation of the physical aspects of the G\"odel universe has led to the introduction of various interesting coordinate systems~\cite{new20,20}; for instance, Gaussian coordinate systems for the G\"odel spacetime have been constructed in Ref.~\cite{new20}.

As in section~\ref{s2}, we are interested in the class of reference
observers that are fixed in space; that is, they have constant $x$,
$y$ and $z$ coordinates. These free observers are endowed with an
orthonormal tetrad frame $\lambda^\mu_{\;\;(\alpha)}$ that is parallel
transported along their geodesic worldlines. Thus
$\lambda^\mu_{\;\;(0)} =u^\mu$ and $\lambda^\mu_{\;\;(i)}$, $i=1,2,3$,
can be expressed as
\begin{align}
\label{eq40} \lambda^\mu_{\;\; (1)} &= \tilde{\lambda}^\mu_{\;\;(1)}
\cos \Omega t+\tilde{\lambda }^\mu _{\;\; (2)} \sin \Omega t,\\
\label{eq41} \lambda^\mu_{\;\;(2)} &=-\tilde{\lambda}^\mu _{\;\;(1)}
\sin \Omega t+\tilde{\lambda }^\mu _{\;\; (2)} \cos \Omega t,
\end{align}
where in $(t,x,y,z)$ coordinates we have
$\tilde{\lambda}^\mu_{\;\;(1)}=\delta^\mu_{\;\; 1}$ and
\begin{equation}
\label{eq42} \tilde{\lambda}^\mu_{\;\;(2)} =\left( -\sqrt{2}
,0,\frac{1}{U(x)},0\right);
\end{equation}
moreover, $\lambda^\mu_{\;\; (3)}=\delta^\mu _{\;\; 3}$ coincides with
the spacelike Killing vector field $\partial_z$. Thus at each point in
space the unit gyro axes of the corresponding reference observer
rotate about the $z$ direction with frequency $\Omega$. For the sake
of simplicity and without any loss in generality, we choose the
reference observer at $x=y=z=0$.

It can be shown that all of the nonzero components of the Riemann
tensor for metric~\eqref{eq37} can be obtained from
\begin{equation}
\label{eq43} R_{0101}  = \Omega^2,\quad R_{0202}=\Omega^2U^2,\quad
R_{0112} =-\sqrt{2}\,\Omega^2 U,\quad R_{1212} =3\Omega^2 U^2,
\end{equation}
via the symmetries of the Riemann tensor. It is then straightforward
to demonstrate that the nonzero components of the projection of the
Riemann tensor on the reference tetrad field can be obtained from
\begin{equation}
\label{eq45} {^F\! R_{0101}} = {^F\! R_{0202}} = {^F\! R_{1212}}
=\Omega^2
\end{equation}
via the symmetries of the Riemann tensor. Using
Eqs.~\eqref{eq2}--\eqref{eq4}, we note that in Fermi coordinates the
spacetime metric---containing only the lowest-order tidal
terms---would be
\begin{equation}\begin{split}
\label{eq46} ds^2 &\approx -[1+\Omega^2 (X^2+Y^2)]
dT^2+dX^2+dY^2+dZ^2\\
&\quad -\frac{1}{3}\Omega^2 (XdY-YdX)^2.
\end{split}\end{equation}

To find the exact expression for this metric, we need to specify all
of the spacelike geodesics of the G\"odel spacetime that are
orthogonal to the worldline of the chosen reference observer
$O:(\bar{t},\mathbf{\bar{x}})=(\tau ,\mathbf{0})$. The
\textit{spacelike} geodesics of the G\"odel spacetime are given by
\begin{align}
\label{eq47} t'+\sqrt{2} Uy'&=E, & \sqrt{2} Ut'+U^2y'&=k,\\
\label{eq48} z'&=h, & {-t'}^2 -2\sqrt{2} Ut'y'+{x'}^2 -U^2{y'}^2+{z'}^2
&=1,
\end{align}
where $t'=dt/d\sigma$, etc. Here $\sigma$ is the proper length of the
spacelike geodesic and $E,k$ and $h$ are constants of integration. We
need a general solution of these equations that would correspond to
the geodesic segment from $Q$ to $P$ in Fig.~1. The vector
tangent to this segment, namely, $(t',x',y',z')$ at $\sigma =0$ is
$\xi^\mu$ at event $Q:(\tau ,\mathbf{0})$; moreover, $\xi^\mu$ is
orthogonal to $\lambda^\mu_{\;\;(0)}$, i.e.
\begin{equation}
\label{eq49} -\xi_\mu \lambda^\mu _{(0)} =t' +\sqrt{2} Uy'=E=0,
\end{equation}
where Eq.~\eqref{eq47} has been used. Therefore, with $E=0$ and
$z=h\sigma$, Eqs.~\eqref{eq47}--\eqref{eq48} imply that
\begin{equation}
\label{eq50} t'=\sqrt{2}\frac{k}{U},\quad {x'}^2 +\frac{k^2}{U^2}
=1-h^2,\quad y'=-\frac{k}{U^2},
\end{equation}
where $h^2\leq 1$. The equation for $x$ can be written as
\begin{equation}
\label{eq51} {U'}^2 -2(1-h^2) \Omega^2U^2=-2\Omega^2k^2,
\end{equation}
which has the general solution
\begin{equation}
\label{eq52} U=\alpha \cosh (a\sigma +b),
\end{equation}
where $a:=\Omega \sqrt{2(1-h^2)}$. Here $\alpha >0$ and $b$ are
constants that are related by the requirement that $U=1$ at $\sigma
=0$, i.e.
\begin{equation}
\label{eq53} \alpha \cosh b=1.
\end{equation}
Substitution of Eq.~\eqref{eq52} in Eq.~\eqref{eq51} results in
\begin{equation}
\label{eq54} \alpha a=\sqrt{2}\Omega |k|.
\end{equation}
It follows from the solution of the other equations in \eqref{eq50}
that
\begin{align}
\label{eq55} t-\tau &= \frac{2k}{\Omega |k|} (\arctan e^{a\sigma +b}
-\arctan e^b),\\
\label{eq56} y&=-\frac{k}{\sqrt{2}\alpha \Omega |k|}[\tanh (a\sigma
  +b)-\tanh b].
\end{align}

Next, we compute $\xi_\mu\lambda^\mu_{\;\;(i)}$ at $Q$, where
$\sigma=0$, and use Eq.~\eqref{eq1} to find Fermi coordinates
$(T,X,Y,Z)$ such that
\begin{align}
\label{eq57}& T=\tau , \quad X\cos \Omega T-Y\sin \Omega T =\sigma
|k|\sinh b,\\
\label{eq58} & X\sin \Omega T+Y\cos \Omega T=-\sigma k,\quad Z=\sigma h.
\end{align}
We note that $X^2+Y^2+Z^2=\sigma ^2$, since $h^2+k^2/\alpha^2=1$. It
proves useful to introduce cylindrical Fermi coordinates $(\rho
,\varphi ,Z)$ such that
\begin{equation}
\label{eq59} X=\rho \cos \varphi , \quad Y=\rho \sin \varphi .
\end{equation}
Then, Eqs.~\eqref{eq57}--\eqref{eq58} can be written as
\begin{equation}
\label{eq60} \sqrt{2} \Omega \rho =a\sigma ,\quad \cos (\varphi
+\Omega T)=\tanh b,\quad \sin (\varphi +\Omega T)=-\frac{k}{|k|}\alpha
.
\end{equation}

Consider now Eq.~\eqref{eq52} for $x:$ expanding its right-hand side
and using Eq.~\eqref{eq60} together with Eq.~\eqref{eq53}, we find
\begin{equation}
\label{eq61} e^{\sqrt{2}\Omega x} =\cosh (\sqrt{2}\Omega \rho )+\sinh
(\sqrt{2} \Omega \rho )\cos (\varphi +\Omega T).
\end{equation}
Similarly, Eq.~\eqref{eq56} for $y$ can be written as
\begin{equation}
\label{eq62} \sqrt{2} \Omega y=\frac{\tanh (\sqrt{2} \Omega \rho )\sin
  (\varphi +\Omega T)}{1+\tanh (\sqrt{2}\Omega \rho )\cos (\varphi
  +\Omega T)},
\end{equation}
while Eq.~\eqref{eq55} for $t-T$ takes the form
\begin{equation}
\label{eq63} \tan \left[ \frac{1}{2} \Omega (T-t)\right]
=\frac{(e^{\sqrt{2}\Omega \rho} -1)\sin (\varphi +\Omega T)}{1-\cos
  (\varphi +\Omega T)+[1+\cos (\varphi +\Omega T)]e^{\sqrt{2}\Omega
  \rho}}.
\end{equation}
It is advantageous to introduce new variables $u$ and $v$ by
\begin{equation}
\label{eq64} u=\sqrt{2}\Omega \rho ,\quad v=\varphi +\Omega T,
\end{equation}
so that the transformation to Fermi coordinates, $(t,x,y,z)\mapsto
(T,X,Y,Z)$, is given by
\begin{align}
\label{eq65} \tan \left[ \frac{1}{2} \Omega (T-t)\right]
&=\frac{(e^u-1)\sin v}{1-\cos v+(1+\cos v)e^u},\\
\label{eq66}e^{\sqrt{2}\Omega x} &=\cosh u+\cos v\sinh u,\\
\label{eq67}\sqrt{2} \Omega y &=\frac{\sinh u\sin v}{\cosh u+\sinh
  u\cos v} ,\quad z=Z.
\end{align}
It should be noted that transformations \eqref{eq66} and \eqref{eq67}
bear a strong resemblance to those used by G\"odel to show the
rotational symmetry of his metric about the $z$ axis (see
Ref.~\cite{17}, p. 449).

One can show that
\begin{align}
\label{eq68} dt+\sqrt{2} U\,dy&=dT+\frac{1}{\Omega} (\cosh u-1)\, dv,\\
\label{eq69} dx^2+U^2\,dy^2&=\frac{1}{2\Omega^2} (du^2+\sinh^2 u\,dv^2),
\end{align}
so that the G\"odel metric
\begin{equation}
\label{eq70} ds^2=-(dt+\sqrt{2}U\,dy)^2 +dx^2+U^2dy^2+dz^2,
\end{equation}
now takes the form
\begin{equation}
\label{eq71} ds^2 =-\left[ dT+\frac{1}{\Omega } (\cosh u-1)\,dv\right]^2
+\frac{1}{2\Omega ^2} (du^2+\sinh^2 u\,dv^2)+dZ^2
\end{equation}
in $(T,u,v,Z)$ coordinates. From $\rho\, d\rho =X\,dX+Y\,dY$ and $\rho^2\,
d\varphi =X\,dY-Y\,dX$, the form of the metric in Fermi coordinates
$(T,X,Y,Z)$ is
\begin{equation}\begin{split}
\label{eq72} ds^2 &= -(1+L)\,dT^2 -2 \Omega FdT(X\,dY-Y\,dX)+dX^2
+dY^2 +dZ^2\\
&\quad +\frac{\mathcal{H}}{\rho^2} (X\,dY-Y\,dX)^2,
\end{split}\end{equation}
where $\rho =\sqrt{X^2+Y^2}\,$, $u=\sqrt{2}\, \Omega \rho$,
\begin{equation}
\label{eq73} L(u) =\frac{1}{2}\sinh^2u,\quad F(u)=\left( \frac{\cosh
  u-1}{u}\right)^2,\quad \mathcal{H}(u)=\left( \frac{\sinh u}{u}\right)^2-1-2 F(u).
\end{equation}
The metric functions $L$, $F$ and $\mathcal{H}$ depend upon $\cosh u$ and $\cosh
(2u)$; expanding these in uniformly convergent power series in
$u=\sqrt{2}\,\Omega \rho$ for any $\rho \geq 0$, we find the complete
tidal expansion of the metric in Fermi coordinates. This result agrees
with Eq.~\eqref{eq46} to order $\Omega^2$.

To examine the admissibility of Fermi coordinates in the sense of
Lichnerowicz~\cite{4}, we need to determine both $(g_{\mu\nu})$ and
$(g^{\mu\nu})$ in Fermi coordinates. Indeed, given our convention that
the signature of the metric is $+2$, Lichnerowicz admissibility
requires that the \textit{principal minors} of these matrices be
negative. We recall that for a symmetric $n\times n$ matrix $M$, the
principal minors are defined by
\begin{equation}
\label{eq75} \det \begin{bmatrix} M_{11}& \cdots & M_{1k}\\
\vdots & & \vdots \\
M_{k1} & \cdots & M_{kk}\end{bmatrix}
\end{equation}
for $k=1,\dots ,n$. Employing $(T,v,\rho ,Z)$ coordinates, it is
possible to show that the admissibility conditions for the inverse
metric are satisfied provided $1+\mathcal{H}$,
\begin{equation}
\label{eq76} 1+\mathcal{H}=\left( \frac{\cosh u-1}{u^2} \right) (3-\cosh u),
\end{equation}
is positive. That is, the admissible Fermi coordinates must remain
within a circular cylinder about the $Z$ axis such that $\cosh u<3$ or
$\rho <\rho_{\max}$, where
\begin{equation}
\label{eq77} \rho_{\max} =\frac{\sqrt{2}}{\Omega}\ln (1+\sqrt{2}
)\approx \frac{1.25}{\Omega }.
\end{equation} 
Expressing the Fermi metric \eqref{eq72} in terms of cylindrical
coordinates $(T,\rho ,\varphi, Z)$, we find
\begin{equation}
\label{eq78} ds^2=-(1+L)\,dT^2-2\Omega \rho^2F\,dTd\varphi+d\rho^2
+\rho^2(1+\mathcal{H})\,d\varphi^2+dZ^2.
\end{equation}
In this form, the G\"odel spacetime region under consideration is
stationary and cylindrically symmetric, and the corresponding Killing
vectors are $\partial_T$, $\partial_\varphi$ and $\partial_Z$. It is
important to note that the square of the magnitude of the azimuthal
Killing vector is $\rho^2(1+\mathcal{H})$, which is positive only for $\rho
<\rho_{\max}$. Thus for a given $T$ and $Z$, the circle of radius
$\rho$ is spacelike within the admissible region, null at $\rho
=\rho_{\max}$ and timelike for $\rho >\rho_{\max}$. The admissible
region therefore excludes closed timelike lines that are known to
exist in G\"odel spacetime~\cite{17,18}.

It is interesting to employ Fermi coordinates to study some of the
properties of G\"odel spacetime. We recall that the source of the
G\"odel solution has four-velocity $u^\mu$ that coincides with the
timelike Killing vector in $x^\mu=(t,x,y,z)$ coordinates of
Eqs.~\eqref{eq37}--\eqref{eq38}. Under the transformation 
$x^\mu\mapsto {\tilde{x}^\mu}=(T,\rho,\varphi ,Z)$, 
$u_\mu \mapsto \tilde{u}_\mu=(\partial x^\alpha /\partial\tilde{x}^\mu)u_\alpha$, 
so that using
$u_\alpha =g_{\alpha 0}$ we have
\begin{equation}
\label{eq79} \tilde{u}_\mu =-\frac{\partial t}{\partial\tilde{x}^\mu}
-\sqrt{2}U(x)\frac{\partial y}{\partial \tilde{x}^\mu} ,
\end{equation}
which can be simply computed using Eq.~\eqref{eq68}. For the metric
$\tilde{g}_{\mu\nu}$ given by Eq.~\eqref{eq78}, $\sqrt{-\tilde{g}}
=(\sqrt{2}\Omega )^{-1} \sinh u$, and the only nonzero components of
$\tilde{g}^{\mu\nu}$ are $\tilde{g}^{\rho\rho}=\tilde{g}^{ZZ}=1$ and
\begin{equation}
\label{eq80} \tilde{g}^{TT} = \frac{\chi -3}{\chi +1} ,\quad
\tilde{g}^{T\varphi}=\tilde{g}^{\varphi T}=-\Omega\, \frac{\chi
  -1}{\chi+1},\quad
\tilde{g}^{\varphi\varphi}=\Omega^2\,\frac{\chi^2+1}{\chi^2-1},
\end{equation}
where $\chi =\cosh u$. It follows that $\tilde{u}^\mu=(1,0,-\Omega
,0)$ in $\tilde{x}^\mu$ coordinates or, expressed geometrically,
$\partial_t=\partial_T-\Omega \partial_\varphi$.

The G\"odel spacetime is of Petrov type D and has five Killing vector
fields. In his original paper~\cite{17}, G\"odel already discussed the
simple symmetries associated with four Killing vectors, which in terms
of $(t,x,y,z)$ coordinates of Eqs.~\eqref{eq37} and \eqref{eq38} are
given by $\partial_t$, $\partial_y$, $\partial_z$ and
$\partial_x-\sqrt{2}\Omega y\partial_ y$. The fifth Killing vector has
a more complicated form
\begin{equation}
\label{eq81} K=2\sqrt{2}U^{-1}\partial_t -2\sqrt{2} \Omega
y\partial_x+(2\Omega ^2y^2-U^{-2})\partial _y.
\end{equation}
This can be derived directly from the Killing equation; alternatively,
one can use the simple Killing vectors in the new forms of the metric
given in Eqs.~\eqref{eq71}, \eqref{eq72} or \eqref{eq78} to find
it. For instance, one can use Eqs.~\eqref{eq65}--\eqref{eq67} to
determine the partial derivatives of $t$, $x$, $y$ and $z$ with
respect to $v$. The result is
\begin{equation}
\label{eq82} \partial_v =\frac{1}{\Omega }(U^{-1} -1)\partial_t
-y\partial _x +\frac{1}{2\sqrt{2}\,\Omega}
 [1+(2 \Omega^2y^2-U^{-2})]\partial_y,
\end{equation}
so that the ``new'' Killing vector $\partial _v$ is expressible in
terms of those in the original coordinate system as
\begin{equation}
\label{eq83} 2 \sqrt{2}\,\Omega \partial_v=-2\sqrt{2}\,
\partial_t +\partial _y+K.
\end{equation}
It is remarkable that pursuing such an approach to the symmetries of
G\"odel spacetime, Obukhov already discovered the coordinate
transformations~\eqref{eq65}--\eqref{eq67} in Ref.~\cite{20}.

\section{Tidal Dynamics in G\"odel Spacetime\label{s5}}

Let us now consider the motion of a free test particle in the exact
Fermi coordinate system that we have constructed in Sec.~\ref{s4}. Our
purpose here is to compare the dynamics in the exact case with that
using only the first-order tidal terms; that is, the solution of the
equations of motion based on the metric form~\eqref{eq46}.

To simplify matters, let us limit our considerations to ``radial''
motion orthogonal to the $Z$ direction. Specifically, we assume that
at $T=0$, $(X,Y,Z)=\mathbf{0}$ and $\mathbf{V}=(V_0,0,0)$, where
$V_0\in [0,1)$. Starting from the exact Fermi metric~\eqref{eq78} in
cylindrical coordinates, we find the equations of motion
\begin{eqnarray}
\label{eq84} \left(\frac{d\rho}{dT}\right)^2&=& \frac{\chi +1}{3-\chi}
-\frac{1}{\Gamma^2_0} \left( \frac{\chi +1}{3-\chi}\right)^2,\\
\label{eq85} \frac{d\varphi}{dT}&=& \Omega\, \frac{\chi-1}{3-\chi},
\end{eqnarray}
where, as in Sec.~\ref{s4}, $\chi=\cosh (\sqrt{2}\Omega \rho)$ and
$\Gamma_0$ is the Lorentz factor corresponding to $V_0$. 
Let $w:=\chi+1$, so that 
\begin{equation}\label{neweq91}
w=2\cosh^2(\frac{\Omega\rho}{\sqrt{2}}).
\end{equation}
For $0\le \rho<\rho_{\max}$, we have that
$2\le w< 4$.
Equation~\eqref{eq84} can be written in terms of $w$ and reduced to quadrature
as 
\begin{equation}\label{neweq92}
\int_2^w \frac{-1+4/x}{\sqrt{(x-2)[4-(2-V_0^2) x]}}\, dx=\pm \sqrt{2}\,\Omega T.
\end{equation}
Next,  formulas 2.261 on p.\ 81 and 2.266 on p.\ 84 of Ref.~\cite{gr} can be employed to express the general solution as 
\begin{equation}\label{neweq84}
-\sqrt{2}\,\arcsin W_1+\frac{1}{\sqrt{2-V_0^2}}\arcsin W_2+\frac{\pi}{2}(\sqrt{2}-\frac{1}{\sqrt{2-V^2_0}})=\pm\sqrt{2}\,\Omega T,
\end{equation}
where $W_1$ and $W_2$ are given by
\begin{eqnarray}
W_1 &=& \frac{(4+V_0^2)-(4-V_0^2)\chi}{V_0^2(\chi+1)},\\
W_2 &=& \frac{2-(2-V_0^2)\chi}{V_0^2}.
\end{eqnarray}
 It follows
from Eq.~\eqref{eq84} that in this ``radial'' motion with zero orbital
angular momentum about the $Z$ axis, $\rho $ increases monotonically
from zero and after reaching a maximum at
\begin{equation}
\label{eq86} \hat{\rho} (V_0)=\frac{1}{\sqrt{2}\Omega} \ln \left(
\frac{\sqrt{2}+V_0}{\sqrt{2}-V_0}\right)
\end{equation}
returns to the origin in a time-symmetric fashion. For $V_0:0\to 1$,
$\hat{\rho}:0\to \rho_{\max}$; that is, a test particle remains within
the Fermi frame, while a null ray reaches the null circular
boundary. An interesting feature of $\ddot{\rho}$, derived from
Eq.~\eqref{eq84}, should be noted: it vanishes at $\rho=0$ for
$V_0=1/\sqrt{2}$, which is the critical speed for relative gravitational motion.
\begin{figure}
\centerline{\psfig{file=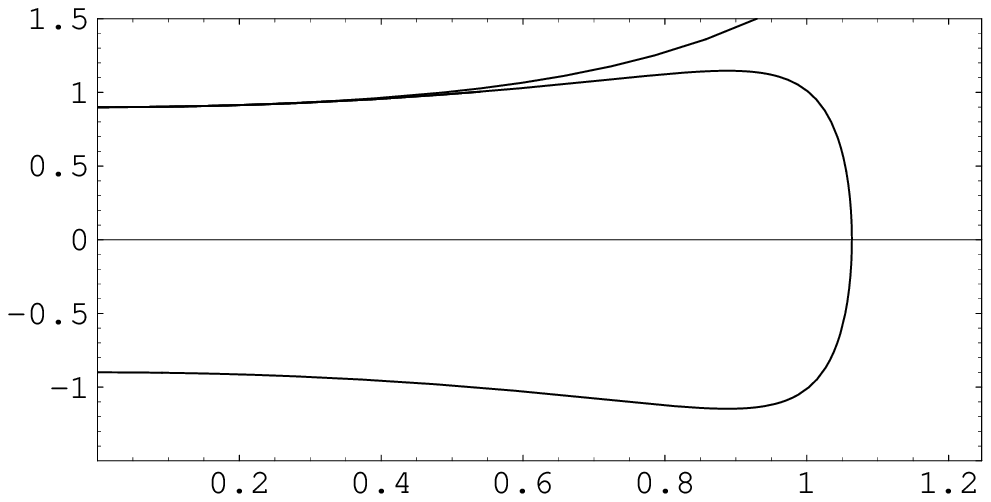, width=15pc}}
\centerline{\psfig{file=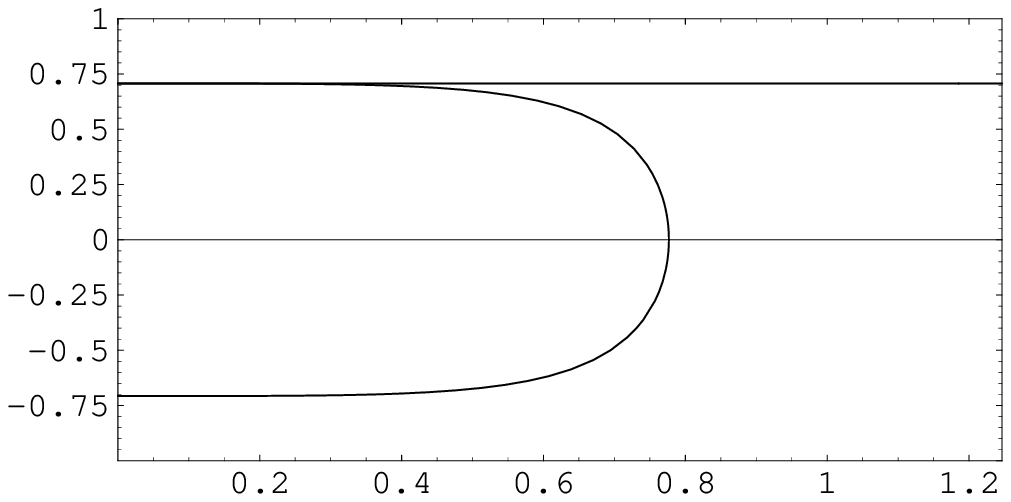, width=15pc}}
\centerline{\psfig{file=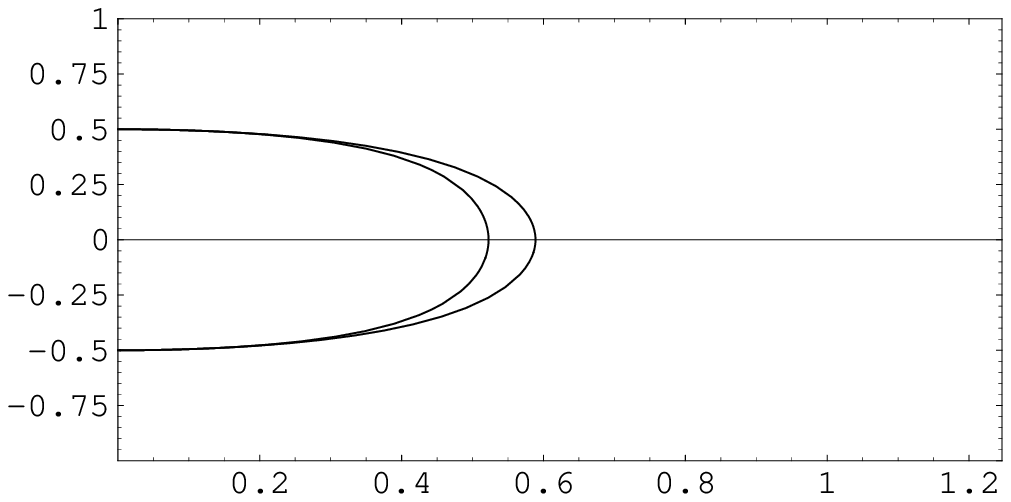, width=15pc}}
\caption{Plot of $\dot \rho$ versus $\Omega\rho$, $0\le\Omega\rho\le \sqrt{2}\ln{(1+\sqrt{2})}$, for $V_0=0.5$, $1/\sqrt{2}$ and $0.9$. In the top and middle panels, the lower (upper) curves represent the motion according to the exact (approximate) equation of motion in Fermi coordinates. In the top panel, $\dot \rho$ monotonically increases in the approximate case.
In the bottom panel, the inner curve corresponds to the exact equation of
motion. 
\label{fig:3}}
\end{figure}

The same kind of motion in the case of first-order tidal terms,
metric~\eqref{eq46}, turns out to be purely radial with the equation
of motion
\begin{equation}
\label{eq87} \ddot{\rho} =-\Omega^2 (1-2\dot{\rho}^2)\rho
\end{equation}
and initial conditions that at $T=0$, $\rho =0$ and
$\dot{\rho}=V_0$.  This equation is expected to be valid for $\Omega \rho$ sufficiently small
compared to unity. 
Equation~\eqref{eq87} can be integrated once and the result is 
\begin{equation}
\label{neweq86} \dot{\rho}^2 =V_c^2-(V_c^2-V_0^2) e^{2\Omega^2 \rho^2};
\end{equation}
moreover, the solution of Eq.~\eqref{neweq86} can be expressed as 
\begin{equation}
\label{neweq87}\Psi_+(\sqrt{2}\Omega \rho; 2V_0^2-1)=\pm \Omega T.
\end{equation}

It is simple to see from the effective potential in Eq.~\eqref{neweq86} that for $V_0>V_c$, the particle monotonically accelerates to almost the local speed of light, while for $V_0=V_c$, the motion is uniform. For $V_0<V_c$, the motion along any radial axis in the $(X,Y)$ plane is periodic and confined to the interval $[-\rho_0,\rho_0]$, where
\begin{equation}
\label{neweq88}\rho_0=\frac{1}{\sqrt{2}\Omega}\sqrt{\ln\Big(\frac{V_c^2}{V_c^2-V_0^2}\Big)}\,.
\end{equation}
We note that for $0<V_0<V_c$, $\rho_0>\hat\rho$, while for $V_0=0$,
$\rho_0=\hat\rho=0$. These results, together with a comparison with the exact solution,  are illustrated in Figure~\ref{fig:3}.  Analytic estimates for the difference between the exact solution and the approximation can be obtained using Eqs.~\eqref{neweq84} and~\eqref{neweq87}.

\section{Discussion\label{s6}}

The  metrics of de Sitter and G\"odel spacetimes in Fermi
coordinates can be expressed in infinite series of tidal terms that
are uniformly convergent over all of space and time; however, the
requirement of (Lichnerowicz) admissibility limits their respective
domains of applicability to $0\leq HR<\pi/2$ and $0\leq \Omega\rho
<\sqrt{2} \ln (1+\sqrt{2})$. In terms of the motion of free test
particles in these domains, it turns out that---for the cases
considered in this paper---the first-order tidal terms already provide
good approximations over significant neighborhoods about the origins
of these domains as illustrated in Figures~2 and~3. It is possible to provide detailed analytic estimates of the difference
between the geodesic (deviation) equation and the generalized Jacobi
equation using the solutions of these equations of relative motion presented
in Secs. III and V.

\appendix
\setcounter{section}{0}
\newcounter{saveeqn}%
\newcommand{\alpheqn}{\setcounter{saveeqn}{\value{equation}}
\stepcounter{saveeqn}\setcounter{equation}{0}
\renewcommand{\theequation}{\mbox{\Alph{section}\arabic{equation}}}}
\renewcommand{\thesection}{\Alph{section}}
\section{Critical Speed}\label{appen:A}
\alpheqn
Imagine the radial motion of a swarm of particles away from a collapsed
configuration as in an astrophysical jet. Relative to a free test observer
and in a Fermi coordinate system based on this observer's worldline, free
particles starting from the position of the observer and moving
ultrarelativistically outward with speed above $V_c = 1/\sqrt{2}$ decelerate
toward the critical speed $1/\sqrt{2}$. However, free particles moving
ultrarelativistically normal to the jet direction accelerate relative to the
observer; via collisions with neighboring particles, the corresponding tidal
energy can be imparted to the collapsed object's environment. One can in
fact envision at the position of the observer a critical velocity cone with
its axis along the jet direction and a total opening angle of $2 \theta_c$ at
its vertex such that $\tan \theta_c = \sqrt{2}$. Within this cone free
ultrarelativistic particles decelerate relative to the observer, while they
accelerate outside the cone. For infrarelativistic motion with speed below
$V_c = 1/\sqrt{2}$, tidal effects tend to behave more or less as one would
generally expect on the basis of Newtonian gravitation theory. That
ultrarelativistic gravitational tidal effects could exhibit astrophysically
interesting features that would be contrary to Newtonian expectations was
first pointed out in Ref.\ \cite{11} and has been the subject of several recent
papers~\cite{12, 13}. It is important to emphasize the general significance of
the critical speed for relative motion in accelerated systems and
gravitational fields. These results follow from the generalized Jacobi
equation that is based on the lowest-order tidal terms; we explain in
Appendix B why it is impractical to use exact tidal terms for these
astrophysically significant problems.

    It is interesting to point out here that in the specific context of
(essentially radial) geodesic motion in the exterior Schwarzschild field,
the critical speed was already discussed by Hilbert~\cite{21}. An examination of
the early references to this subject is contained in Ref.~\cite{22}. In terms of
Schwarzschild coordinates, the critical speed is given by $v_c = 1/\sqrt{3}$ as
discussed in detail in Ref.~\cite{23}. Using a more invariant approach, McVittie
\cite{24} derived the critical speed $V_c = 1/\sqrt{2}$ for geodesic motion in Schwarzschild spacetime.

    Shapiro's observation of the gravitational time delay can be interpreted
to imply that light slows down in the gravitational field of a massive body.
This is in apparent conflict with the fact that in Newtonian gravity
particles speed up as they fall toward a massive object. These ideas are
properly integrated in general relativity through the concept of critical
speed. In this context, $v_c$ and $V_c$ have been treated in Ref. \cite{23}, which
should be consulted for a more detailed treatment of (and further
references to) this topic. Finally, we note that the critical speed $v_c =
1/\sqrt{3}$ has been discussed recently in connection with the deflection of
particles by a radially moving gravitational lens~\cite{25}.

\section{Fermi Coordinates in Schwarzschild Spacetime}\label{appen:B}
The purpose of this appendix is to bring out the difficulties 
encountered in attempts to employ \emph{explicit} Fermi coordinates in
black-hole spacetimes. To this end, we imagine in the following treatment
the simplest situation of astrophysical interest, namely, purely radial motion in
the exterior Schwarzschild gravitational field.

Consider the exterior Schwarzschild spacetime of a spherical source of mass $M$ represented by the metric
\begin{equation}\label{B1}
ds^2=-\Big(1-\frac{2M}{r}\Big)\,dt^2+\Big (1-\frac{2M}{r}\Big )^{-1}\, dr^2+r^2 (d\theta^2+\sin^2\theta\, d\phi^2)
\end{equation}
for $r>2M$. To simplify matters, we limit our treatment to the motion of test particles along a fixed radial direction, which we can choose   to be the $z$ axis (i.e. $\theta=0$) with no loss of generality.

To construct Fermi coordinates for this two-dimensional world, we must specify a reference observer.
Imagine, therefore, a free observer $O:(\bar t,\bar r)$ that starts from $r_0>2M$ at $\bar t=0$ and follows an escape trajectory that reaches radial infinity with zero speed. The geodesic equations of motion for this observer are
\begin{equation}\label{B2}
\frac{d\bar t}{d\tau}=\frac{1}{1-\frac{2M}{\bar r}},\qquad
\frac{d\bar r}{d\tau}=\sqrt{\frac{2M}{\bar r}},
 \end{equation}
where $\tau$ is the observer's proper time such that $\tau=0$ at $\bar t=0$ and $\bar r=r_0$. The system~\eqref{B2} can be integrated and the result is
\begin{eqnarray}
\label{B3} \bar r=2 M\cosh^2 w, \qquad 3\tau -4 M \cosh^3 w=c_1,\qquad{}\\
\label{B4} 3\bar t-4 M \cosh^3w-12 M( \cosh w+\ln \tanh \frac{w}{2})=c_2,
\end{eqnarray}
where $c_1$ and $c_2$ are constants of integration and can be expressed in terms of $r_0$ using the initial conditions.

The observer carries an orthonormal parallel-propagated tetrad $\lambda^\mu_{\;\;(\alpha)}$ along its path. This local frame is given by unit vectors along the temporal and radial directions
\begin{eqnarray}
\label{B5} \lambda^\mu_{\;\;(0)}=((1-\frac{2M}{\bar r})^{-1}, \sqrt{\frac{2M}{\bar r}}\,, 0,0) ,\\
\label{B6} \lambda^\mu_{\;\;(3)}=(\sqrt{\frac{2M}{\bar r}}\,(1-\frac{2M}{\bar r})^{-1},1, 0,0),
\end{eqnarray}
in  $(t, r,\theta,\phi)$ coordinates, respectively, as well as $\lambda^\mu_{\;\;(1)}$ and $\lambda^\mu_{\;\;(2)}$.
The axial symmetry of the spacetime about the $z$ axis implies that a rotational degeneracy exists in the choice of $\lambda^\mu_{\;\;(1)}$ and $\lambda^\mu_{\;\;(2)}$. But, once these unit vectors are chosen at $\bar t=0$, they are then parallel transported along the path of $O$. 

Let us next consider the spacelike geodesics appropriate to the two-dimensional world of radial motion under consideration. They are given by
\begin{equation}\label{B7}
\frac{d t}{d\sigma}=\frac{\hat p}{1-\frac{2M}{r}},\qquad
\frac{d r}{d\sigma}=\epsilon \sqrt{\hat p^2 +1 -\frac{2M}{r}},
\end{equation}
where $\hat p$ is an integration constant and $\epsilon=\pm 1$. As in Figure~\ref{fig:1}, we need only those spacelike geodesics that are orthogonal to the reference worldline at $Q$. It follows from $\xi_\mu \lambda^\mu_{\;\;(0)}=0$ that 
\begin{equation}\label{B8}
\hat p=\epsilon \sqrt{\frac{2M}{\bar r}}\,.
\end{equation}
Moreover, it follows from Eq.~\eqref{eq1} that in this case the Fermi coordinates are
\begin{equation}\label{B9}
T=\tau,\quad X=Y=0,\quad Z=\epsilon \sigma.
\end{equation}
It remains to integrate system~\eqref{B7} from $Q$ to $P$ for constant $T$. This system can be written as 
\begin{equation}\label{B10}
\frac{d t}{dZ}=\frac{ p}{1-\frac{2M}{r}},\qquad
\frac{d r}{dZ}= \sqrt{q^2  -\frac{2M}{ r}},
\end{equation}
where $p=\epsilon \hat p$ and $q$ are functions of $T$, 
\begin{equation}\label{B11}
 p=\sqrt{\frac{2M}{\bar r}},\qquad
q= \sqrt{1 +\frac{2M}{\bar r}}.
\end{equation}
The solution of system~\eqref{B10} is simplified if we introduce a new quantity $\zeta$ such that
\begin{equation}\label{B12}
\sqrt{\frac{r}{2M}}=\frac{1}{q} \cosh \zeta.
\end{equation}
Then, we find
\begin{eqnarray}
\label{B13}\zeta+\frac{1}{2} \sinh (2\zeta)-\bar \zeta-\frac{1}{2} \sinh (2\bar \zeta)= \frac{q^3}{2M} Z,\\
\label{B14} \mathcal{F}(\zeta)-\mathcal{F}(\bar\zeta)+\frac{2p}{q}
(\zeta-\bar\zeta)=\frac{1}{2M}[t-\bar t(T)-p Z], 
\end{eqnarray}
where $\bar\zeta$ is given by 
\begin{equation}\label{B15}
\bar\zeta=\cosh^{-1}(\frac{q}{p})
\end{equation}
and $\mathcal{F}(\zeta)$ is defined  to be
\begin{equation}\label{B16}
\mathcal{F}(\zeta)=\ln(\frac{e^{2\zeta}-A}{e^{2\zeta}-B}).
\end{equation}
Here $A$ and $B$ are functions of $T$ given by
\begin{equation}\label{B17}
A=1+2p^2+2pq,\qquad B=1+2p^2-2pq.
\end{equation}

Inspection of Eq.~\eqref{B13} makes it evident that it is not possible to
express $\zeta$ \emph{explicitly} in terms of $T$ and $Z$. It follows that the radial
coordinate $r$ in the Schwarzschild metric~\eqref{B1} cannot be expressed
explicitly in terms of $T$ and $Z$. For the explicit expression of the metric in
Fermi coordinates, one must resort to expansions in powers of $Z$. It is therefore clear that the transformation $(t,r)\mapsto (T,Z)$ is given only \emph{implicitly} by Eqs.~\eqref{B12}--\eqref{B17}. Taylor expansions can be used to obtain an explicit form for the transformation to Fermi coordinates. 
We find that 
\begin{eqnarray}
\label{B18}t&=&\bar t(T)+f_1Z+\frac{1}{2!}f_2 Z^2+\frac{1}{3!} f_3 Z^3+\mathcal{O}(Z^4),\\
\label{B19} r&=&\bar r(T)+Z+\frac{1}{2}\frac{M}{\bar r^2} Z^2-\frac{1}{3} \frac{M}{\bar r^3} Z^3+\mathcal{O}(Z^4), 
\end{eqnarray}
where
\begin{equation}\label{B20}
f_1=\frac{p}{1-p^2},\quad f_2=-\frac{1}{\bar r} \frac{p^3}{(1-p^2)^2},\quad f_3=\frac{1}{2\bar r^2}\frac{p^3(4-p^2+p^4)}{(1-p^2)^3}.
\end{equation}
Using transformations~\eqref{B18}--\eqref{B19}, we find the relevant metric coefficients in Fermi coordinates
\begin{equation}\label{B21}
g_{TT}=-1+\frac{2M}{\bar r^3} Z^2+\mathcal{O}(Z^3),\quad g_{TZ}=g_{ZT}=\mathcal{O}(Z^3),\quad g_{ZZ}=1+\mathcal{O}(Z^3),
\end{equation}
in agreement with Eqs.~\eqref{eq2}--\eqref{eq4}, since in this case
\begin{equation}\label{B22}
R_{TZTZ}=-\frac{2M}{\bar r^3},\quad \bar r=(r_0^{3/2}+\frac{3}{2}\sqrt{2M}\, T)^{2/3}.
\end{equation}

 We have shown that it is not possible to provide explicit Fermi
coordinates for radial motion in the exterior Schwarzschild spacetime. The
upshot of our direct approach is just a different approximation scheme for
determining the series of tidal terms in Eqs.~\eqref{eq2}--\eqref{eq4}. In this way, we have also provided the justification for using the generalized Jacobi equation in our previous work~\cite{11,12,13}.

\section{Extent of validity of the generalized Jacobi approximation}\label{appen:C}
The generalized Jacobi equation~\eqref{eq11} is a first-order approximation to
the geodesic deviation equation~\eqref{eq8} in Fermi normal coordinates. The purpose of this appendix is to describe a general method for determining the
timescale over which the first-order approximation is valid.

It is convenient to express Eqs.~\eqref{eq8} and~\eqref{eq11} in dimensionless form; to
this end, we introduce $\mathcal{T}:= T/\mathcal{R}$ and $\mathbf{\mathcal{X}} := \mathbf{X}/\mathcal{R}$ ,
where $\mathcal{R}$ is a constant effective radius of curvature of the spacetime
under consideration such that $|\mathbf{\mathcal{X}}| < 1$ in the domain of
admissibility of Fermi coordinates. In  terms of $\mathcal{U} :=  (\mathbf{\mathcal{X}}, \mathbf{V})$, Eqs.~\eqref{eq8} and~\eqref{eq11}  can be expressed respectively as
\begin{eqnarray}\label{C1}
   \mathcal{U}' &=& (\mathbf{V}, \mathbf{P}(\mathcal{U}) + \mathbf{Q}(\mathcal{U})),\\     
\label{C2}\tilde{ \mathcal{U}}' &=& (\tilde{\mathbf{V}}, \mathbf{P}(\mathcal{\tilde{U}})),             
\end{eqnarray}
where a prime denotes differentiation with respect to $\mathcal{T}$. Here
\begin{eqnarray}
\nonumber    P^i(\mathbf{\mathcal{X}}, \mathbf{V})&:=&-\mathcal{R}^2 [{^F\! R_{0i0j}}+2 {^F\! R_{ikj0}}V^k\\
\label{C3} &&{}+\frac{2}{3} (3 {^F\! R_{0kj0}} V^iV^k +
    {^F\! R_{ikjl}}V^kV^l +{^F\! R_{0kjl}}V^iV^kV^l)]\mathcal{X}^j , \\        
\label{C4}     Q^i(\mathbf{\mathcal{X}}, \mathbf{V}) &:=& - [\mathcal{R}(\Gamma^i_{\alpha\beta}-\Gamma^0_{\alpha
  \beta}V^i)V^\alpha V^\beta+P^i(\mathbf{\mathcal{X}}, \mathbf{V})] ,                  
\end{eqnarray}
where $V^\alpha := dX^\alpha/dT = (1, \mathbf{V})$.

To estimate the difference between the solutions of Eqs.~\eqref{eq8} and~\eqref{eq11}, we
introduce $\mathcal{Z} := \mathcal{U} - \mathcal{\tilde{U}}$ and note that
\begin{equation}\label{C5}
     \mathcal{Z}' = J (\mathcal{U}) - J (\tilde{\mathcal{U}}) + S (\mathcal{U}),   \end{equation}
where
\begin{equation}\label{C6}
J(\mathcal{U}) := (\mathbf{V}, \mathbf{P}(\mathcal{U})),
\qquad    S(\mathcal{U}) := (0, \mathbf{Q}(\mathcal{U})).
\end{equation}

The initial conditions for Eqs.~\eqref{eq8} and~\eqref{eq11} are the same, hence it follows from the integration of Eq.~\eqref{C5} that
\begin{equation}\label{C7}
|\mathcal{Z}(\mathcal{T})| \le \int_ 0 ^ \mathcal{T}  |J(\mathcal{U})  - J(\tilde{\mathcal{U}})|\, d\mathcal{T}' + \int_ 0 ^ \mathcal{T} |S(\mathcal{U}) |\, d\mathcal{T}'.  
\end{equation}
It is clear from definitions~\eqref{C3} and~\eqref{C4} that $\mathbf{Q}$ consists of tidal terms of second, third and higher orders. Since $|\mathbf{\mathcal{X}}|<1$ and
$|\mathbf{V}|\lesssim 1$, we may suppose that $\|DJ\|\le  A_0$ and $\|S\|\le B_0$ over an interval of interest $[0, \mathcal{T}]$ for some positive constants $A_0$ and $B_0$.
Hence it follows from Eq.~\eqref{C7}  that
\begin{equation}\label{C8}
   |\mathcal{Z}(\mathcal{T})|\le A_0 \int_0^\mathcal{T} |\mathcal{Z}(\mathcal{T}')|\,d\mathcal{T}' + B_0 \mathcal{T}.     
\end{equation}
By an application of Gronwall's inequality~\cite{ccc}, we have the fundamental
estimate
\begin{equation}\label{C9}
   |\mathcal{Z} (\mathcal{T})| \le B_0 \mathcal{T} e^ {A_0 \mathcal{T}} ,  
\end{equation}
which implies that
\begin{equation}\label{C10}
      |\mathbf{\mathcal{X}} (\mathcal{T}) - \mathbf{\tilde{\mathcal{X}}} (\mathcal{T})| \le  B_0 \mathcal{T} e^ {A_0\mathcal{T}} . 
\end{equation} 

    As a measure of the validity of the approximation of $\mathbf{\mathcal{X}} (\mathcal{T})$
by $\mathbf{\tilde{\mathcal{X}}}(\mathcal{T})$, we can determine the timescale $\mathcal{T}_\epsilon$
over which the magnitude of their difference is less than $\epsilon$, $0<
\epsilon \ll 1$; that is,
\begin{equation}\label{C11}
    B_0 \mathcal{T}_\epsilon  e^ {A_0 \mathcal{T}_\epsilon} = \epsilon.   
\end{equation}
Using Lambert's $W$ function, which is the inverse of the function $x \mapsto x
e^x$, we find
\begin{equation}\label{C12}
   \mathcal{T}_\epsilon = \frac{1}{A_0} W \Big(\frac{ A_0 \epsilon}{B_0}\Big).  
\end{equation}
It should be clear that this method provides a general but crude estimate;
indeed, sharper estimates may be obtained in specific cases using explicit
solutions as in Secs.~\ref{s3} and~\ref{s5}  of this paper.

\acknowledgments{BM is grateful to Valeri Frolov for interesting discussions.}

\end{document}